\documentclass[aip]{revtex4-1}
\usepackage{graphicx}

\def\half {{1\over 2}}

\def\bea{\begin{eqnarray}}
\def\eea{\end{eqnarray}}

\def\sqr#1#2{{\vcenter{\vbox{\hrule height.#2pt
      \hbox{\vrule width.#2pt height#1pt \kern#1pt
         \vrule width.#2pt}
      \hrule height.#2pt}}}}

\def\figloc#1#2 {
\begin{figure}\begin{center}
    \includegraphics[width=80mm]{#1}
    \caption{ #2}
    \end{center}\end{figure}
}

\begin{document}
\title{Kepler's Laws without Calculus  }

\author{W. G. Unruh}
\affiliation{ CIfAR Cosmology and Gravity Program\\
Dept. of Physics\\
University of B. C.\\
Vancouver, Canada V6T 1Z1\\
~
email: unruh@physics.ubc.ca}

~

~

\begin{abstract}
	Kepler's Laws are derived from the inverse squared force law without
	the use of calculus and are  simplified over
	previous such derivations. It uses just elementary algebra and
	trigonometry, and does not even use any advanced geometry.

\end{abstract}

\maketitle
\section{}
Newton in the Principia used the inverse squared force law  (and Galileo's idea of compound
motion) to derive Kepler's laws. As usual for him, the proof is a purely
geometric proof, using no calculus, although he might well have used calculus
in private. Whether or not his proof that inverse-square implied conic section
as the only solutions has been the topic of some controversy\cite{weinstock}
but that is not the topic of this paper. Maxwell \cite{maxwell}, using the
hodographic technique of Hamilton\cite{hamilton,derbes}  gave a very
different proof. In the 1960's,  Feynman\cite{feynman} gave a geometric proof
very similar to Maxwell's.
Finally, Vogt\cite{vogt} in the American Journal of Physics also carried out a
derivation which started out with the energy conservation equation and the
angular momentum conservation to again present a geometric proof.

In all of these cases, the derivation that the orbit actually is an ellipse
was, to me,  somewhat  torturous and difficult to follow. 
In the following, starting off using the Maxwell-Feynman technique of solving
for the velocities as a function of the angle about the center of attraction,
the
derivation that the orbit is an ellipse is simplified about as much as
possible. I do so both  using calculus and, following Newton's proof of
Kepler's second law, a discrete geometric argument to arrive at the equation
for a conic section.

\section{Kepler's second law}

In keeping with all of the above, I will start with Newton's discretized proof
of Kepler's second law,
showing that a central force produces orbits which obey the equal areas in
equal times law. 
I will repeat it because I will use the same ideas in the proof of the first law. 

\figloc{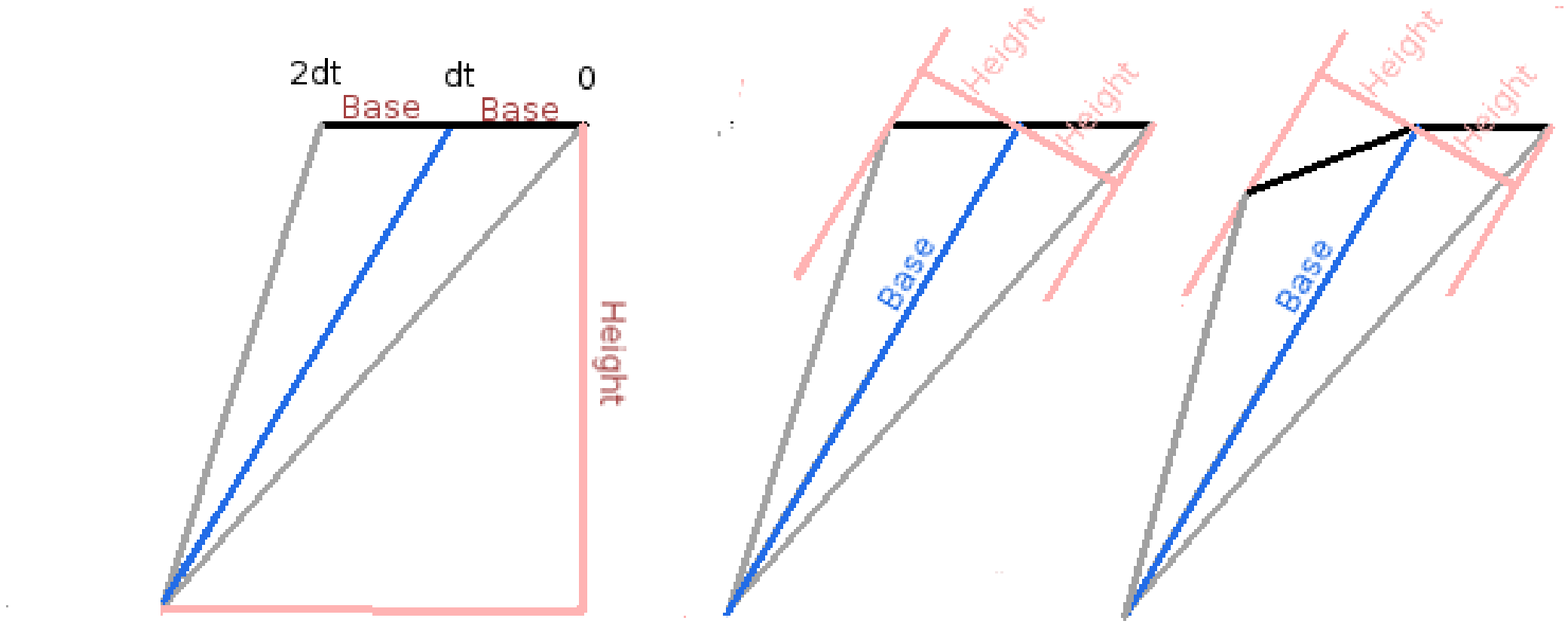}{These three diagrams show the two triangular areas
	swept out in equal times by a polygonal orbit with a central force
	acting between two situations in which the object travels inertially.
	The first figure shows that if there is no force acting at all, the
	areas of the two triangles are equal since the bases are of equal
	length since the speeds are the same. The second figure shows that the
	perpendicular to the common side to the height of the end points of
	the  line must therefor also be the same. In the third figure that is
	an acceleration acting along the common line, and perpendicular to the
orthogonal line to this common line.}

Consider first a straight line along which an object travels inertially. This
means it continues in the same direction with constant speed. Consider the
motion in two adjacent very short time intervals, of length $dt$ where this is
assumed small. Join the points at times $0,dt,2dt$ to what will be the centre
of force. We now have two triangles, each with base of length $|v|dt$ and
common vertex. They must therefor have the same area. Now look at these
triangles and the common base joining the the point at time $dt$ to the vertex.
Since the areas are common, the heights to the tips at $0$ an $2dt $ must be
equal. Draw the perpendicular to the the line from the vertex to the point
$dt$  to the two lines which go through those two tips and are parallel to
line from the vertex to $dt$. The two line segments from $dt$ to the parallel
through $0$ and from $dt$ to the parallel through $2dt$ are equal as these are
just the heights of the triangles from the common base. Ie, the particle
travels with uniform speed along this straight line. Galileo stated that the
if there was an acceleration perpendicular to this line, that uniform motion
along the line would not be disturbed. Now accelerate the body suddenly along
the radius through $dt$. The second triangle will be distorted, but its common
base will not change, and their  heights will also not change. Thus the areas
of the triangles will not change. If we now take the limits as $dt$ goes to
zero, we find that the areas swept in equal times will be constant no matter
what the magnitude is of the radial force. This is just the expression that a
radial force has a conservation law, the conservation of angular momentum. 

We can write the above in terms of differentiation In the limit as $\Delta
theta$ goes to 0, the area is $ r^2 ({d\theta/ dt})/2={\ell/2}$ where
$\ell$ is thus twice the rate of change of area with time and is equal to the
angular momentum. 
\bea
{d\theta\over dt}={\ell\over r^2}.
\eea

\section{Inverse square law and Ellipse--Calculus}

Let me first go through the derivation using calculus, but not solving any
differential equations. Since the geometric proof will follow the procedure in
detail, I will first solve the equations using calculus.

Consider that the force law is an inverse square force law. Ie, the
acceleration of the object has the form $\mu  / r^2$ where $r$ is the
distance from the object to the above vertex. Now consider the change in
velocity, but not in a unit time but in a unit angle.  We can write the
equation of motion in a small time interval as

In differential form, Newton's second law becomes, 
\bea
{d\vec v\over d\theta}= {2r^2\over \ell} (-{\mu \over r^2} \vec n_r)
\eea
from which 
\bea
\vec v=\vec v_0 + {\mu \over \ell} \vec n_\theta
\eea
Where $r\vec n_r= {\vec r}$ and $\vec n_\theta$ is the unit vector orthogonal to $\vec
n_r$ and points in the direction of increasing $\theta$.

\bea
{1\over r}{d(r \vec n_r)\over d\theta}={r\over \ell}\left(\vec v_0 +{\mu \over
l}\vec n_\theta\right)
\eea

Taking only the tangential component, we have
\bea
1&=& {r\over \ell} (\vec v_0\cdot \vec n_\theta) +{2\mu \over \ell})\\
 &=& {|v_0|\over \ell} r \cos(\theta) +{\mu \over \ell^2} r
\eea
or 
\bea
&&r+ {|v_0|\ell\over\mu }x={\ell^2\over \mu } \\
\eea
 with $x=r\cos(\theta)$, or
\bea
r+\epsilon x = R_0
\eea
where $\epsilon= {|v_0|\ell/\mu }$ and $R_0= {\ell^2/ \mu }$.

This is the equation for an ellipse with  the semi-major axis  $$a={R_0/
(1-\epsilon^2)}= {({l^2\mu }/ \mu^2)-\left({|v_0|l }\right)^2}$$.

\section{Ellipse with Geometry}
Now, let us try doing the above using geometry rather than calculus. (It is
suspected that Newton solved many problems using calculus that he had
invented, and translated them into geometry for publication.) This provides a
way of proving that that orbit is elliptical without calculus. 

\figloc{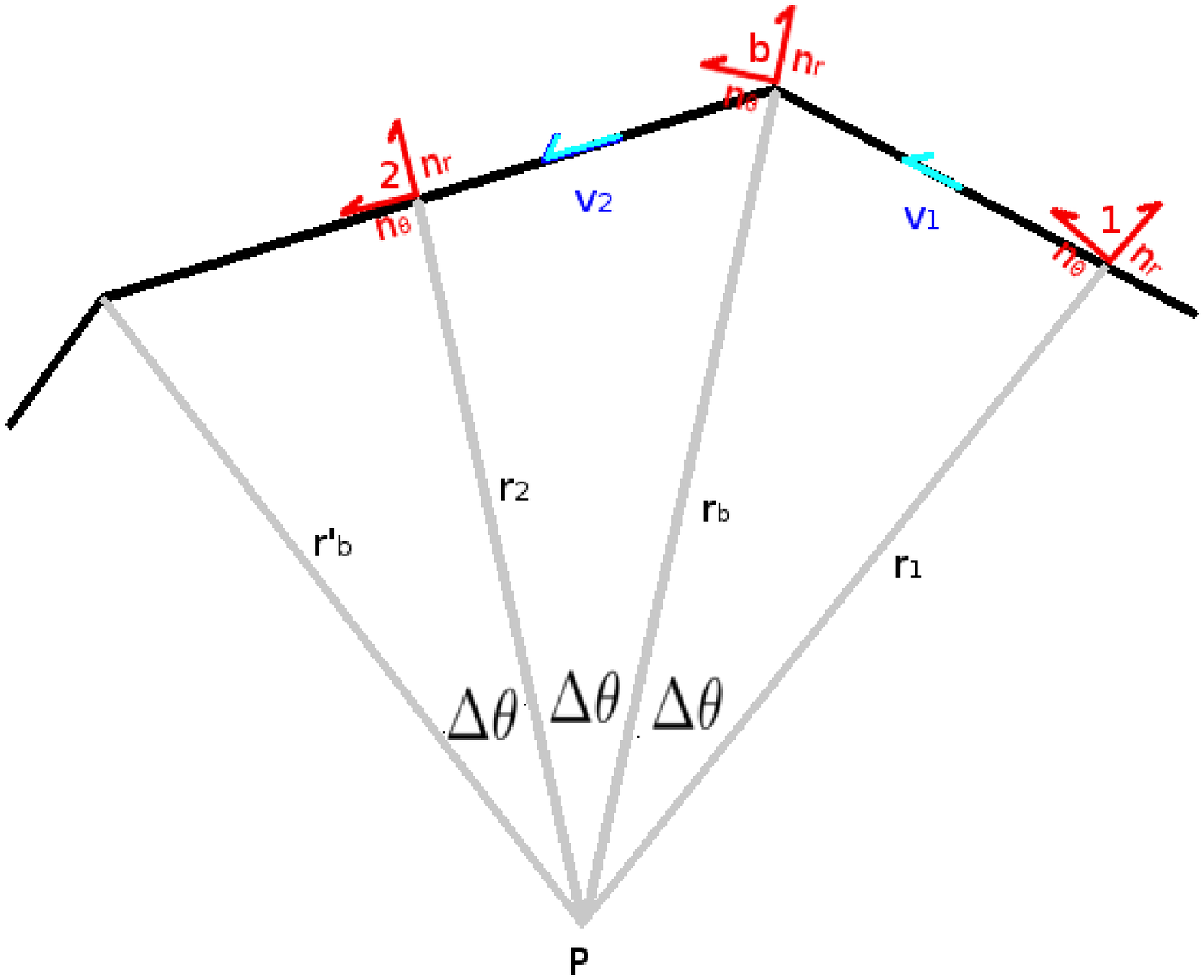}{The diagram of the polygonal approximation to the
	orbit, with the force acting instantaneously at the points b and b'.
}

We follow the spirit of the procedure used to prove the second law, but now
look at a larger section of the obit. We again draw the orbit as a polygon but
now with the various lines defining triangles from the vertex P being separated
by angles $\Delta\theta$. Every second radial line goes to a vertex where the
velocity changes discontinuously as in the first law proof. The other radial
lines go from the vertex to a point near the midpoints of the lines between the
discontinuity vertexes. Each pair of adjacent radial lines are separated by
the above angle. The area of the triangle from the centre P to point b to
point 1 is given by \bea
A_1= \half r_b  r_1 \sin(\Delta \theta) 
\eea
while the second triangle is 
\bea
A_2=\half r_b r_2 \sin(\Delta\theta)
\eea
The  change in velocity at the vertex $b$ is $\Delta\vec v=\vec v_2-\vec v_1$. It is
assumed that there is a sudden acceleration $\vec a$ along $r_b$. Newton's
equation gives
\bea
\delta \vec v=  \vec {\bf a}  T_{12}
\eea
where $ T_{12}$ is the time of travel between points 1 and 2 and $\delta
Q=Q_2-Q_1$ for any quantity $Q$. But by Kepler's
second law, 
\bea
 T_{12}= {2\over \ell} (A1+A2)
\eea
where $\ell$ is again is twice the rate at which the area is swept out by the radius vector.
Thus we have
\bea
\delta \vec v= \vec{\bf a} {1\over \ell} (r_1+r_2)r_b \sin(\Delta \theta)
\label{hom}
\eea
where $\vec {\bf a}$ is the acceleration vector.
Again use the unit vectors $\vec n_r,~\vec n_\theta$.

Assume that
\bea
\half (r1+r2)r_b \vec {\bf a}=-\mu \vec n_{rb}
\eea
where $\mu $ is a constant.  This is the assumption that the acceleration
falls off as the inverse of the distance squared (Newton's law of
gravitation) adapted to the discretized version. 
$(r_1+r_2)r_b)$ is of order $\Delta\theta^2$ away from  $r^2$ 

Using that the vectors $ \vec n_{r1} $ and $\vec n_{\theta1}$ are just
rotations through the angle $-\Delta \theta$ from those at the point $b$, and
similarly for the vectors at point 2 through the angle $\Delta\theta$, one has
\bea
\vec n_{r1} +\vec n_{r2}= 2 \cos(\Delta \theta) \vec n_{rb}\\
\vec n_{\theta 1} +\vec n_{\theta 2}= 2 \cos(\Delta \theta) \vec n_{\theta b}\\
\vec n_{r2} -\vec n_{r1}= 2 \sin(\Delta\theta) \vec n_{\theta b}\\
\vec n_{\theta 2} -\vec n_{\theta 1}= 2 \sin(\Delta\theta) \vec n_{r b}
\eea

We can write 
\bea
\vec v= v_r \vec n_r +v_\theta \vec n_\theta
\eea
and the equation for the discontinuity of $\vec v$ caused by the acceleration
at $b$ is 
\bea
\half \left[\delta v_{r} \overline{\vec n_r} +\overline{ v_{r}} \delta{\vec n_r} 
+\delta v_{\theta } \overline{\vec n_{\theta}} +\overline{ v_{\theta }}
\delta{\vec n_r}\right]
=- {\mu \over \ell} \vec n_{rb}\sin(\Delta \theta) 
\eea
where $\delta Q=Q_2-Q_1$ and $\overline Q= (Q2+Q1)$ for any quantity $Q$.

Using the equations for the sums and differences and separating out the
components of $\vec n_{rb} $ and $\vec n_{\theta b}$, we have the two
equations for the components
\bea
2\delta v_r \cos(\Delta\theta) - 2\overline {v_\theta}\sin(\Delta\theta)= {2\mu \sin(\Delta\theta)\over \ell} \\
2\delta v_\theta \cos(\Delta\theta) +2\overline {v_r}\sin(\Delta \theta) =0
\eea
which are a pair of difference equations for $v_{r 2}$ and $v_{\theta 2}$ in
terms of $v_{r 1}$,  $v_{\theta 1}$ and $\mu$.

The solution of these linear equations can always be written as a sum of a particular
solution of the inhomogeneous equations plus an arbitrary solution of the homogeneous
equations ($\mu=0$). The simplest solution to the former is to take 
$ v_{ri}=0$ for all numerical $i$. Then $v_{\theta i}$ are all equal,
$\hat v_{\theta}$.
One has
\bea
2\hat v_{\theta}\sin(\Delta\theta)= {2\mu \sin(\Delta\theta)\over \ell}\\
\eea
or
\bea
\hat v_{\theta}= {\mu\over \ell}
\eea

The simplest way to solve the homogeneous equation is to go back to eqn
\ref{hom} setting the rhs (proportional to $\mu$)  to zero, which gives 
\bea
\vec v_2=\vec v_1
\eea
The homogeneous solution is an arbitrary constant
vector $\vec v_0$ independent of which vertex one is looking at.

The generic solution is exactly the solution obtained above for the
calculus solution, namely a constant vector plus a ``rotating" vector
orthogonal to the radius vector. 

Now let us solve for the equation of the polygon. The vectors $\vec v_0 + \hat
v_{\theta i} \vec n_{\theta i}$ point along the sides of the polygon and the vector
difference of the two location vectors $\vec r'_b$ and $\vec r_b$ is just the
velocity  $\vec v_2$ times the time $T_{bb'}$ it takes the particle to travel
between the two points $b$ and $b'$.

\bea
r'_b\vec n'_{rb}-r_2 \vec n_{rb}= (\vec v_0 +{\mu\over l}\vec n_{\theta
2}) T_{2b'}\\
\eea
or, using Kepler's second law with $T_{2b'}=[r'_b \sin(\Delta\theta)]/2\ell$ as
the time to go from the point 2 to b'. The equation for $r_b\vec n_{rb}-r_2
\vec n_{rb}$ is the same with $\Delta\theta\rightarrow -\Delta\theta$  We now take the dot product of this
with $\vec n_{\theta2}$
Taking the dot product with $\vec n_{\theta 2}$  and dividing by $r_b$ we get
\bea
1= \left(\vec n_{\theta 2}\cdot \vec v_0 +{\mu\over \ell}\right) {r_2 \over \ell}
\eea
which, after defining $n_{\theta 2}\cdot \vec v_0=|v_0|\cos(\theta_2) $ and $r_2
\cos(\theta_2)=x_2$, gives
\bea
r_2+{|v_0|\ell\over \mu}x_2= {\ell^2\over \mu}
\eea
which is again exactly the equation for an ellipse passing through the
endpoints of lines from the vertex P to the straight segments of the polygon,
near the centres of those lines. 
This is true even in the limit as $\Delta \theta$ goes to zero giving exactly
the same solution as above done with calculus. This analysis is identical to
that using $r_b$ since the only difference was taking $\Delta\theta\rightarrow
-\delta\theta$.

We now take the dot product with $\vec n_{r2}$ and get
\bea
r_b' \cos(\Delta\theta) - {R_0\over 1+\epsilon \cos(\theta_2)}= ({v_0\over
\ell} \sin(\theta_2) ) rb'r2
\sin(\Delta\theta)= {\epsilon \sin(\Delta\theta) \over 1+\epsilon \cos(\theta_2)
}r_b
\eea
or solving for $r_b'$, 
\bea
r_b'{\left( \cos(\Delta\theta) +\epsilon(\cos(\theta_2) \cos(\Delta\theta) -
\sin(\theta_2)\sin(\Delta\theta)\right)\over 1+\epsilon \cos(\theta_2)} = r_2
\eea
or
\bea
r'_b= {R_0\over \cos(\Delta\theta) +\epsilon \cos(\theta_2+\Delta\theta)}
\eea
where $\cos(\theta_2+\Delta\theta)=\cos(\theta_{b'})$.

One gets exactly the same equation for $r_b$ but with $\Delta\theta\rightarrow
-\Delta\theta$ and thus in terms of $\theta_b$.

Thus all of the $b$ vertexes lie on the ellipse 
\bea
r +{\epsilon\over \cos(\Delta\theta)} x = {R_0\over \cos(\Delta\theta)}
\eea

Thus we can exactly solve the equations of motion for a discretization of the
orbit, a la Newton's proof of the second law without doing any calculus
whatsoever (but with some algebra and trigonometry). In particular, there is
no need to solve any differential equation. It also does not require any
tricky geometric proofs, since geometry has, unfortunately, been removed from
many high school and college curricula.  This makes this proof that Kepler's first law is obeyed accessible to
even good high school students. 

\figloc{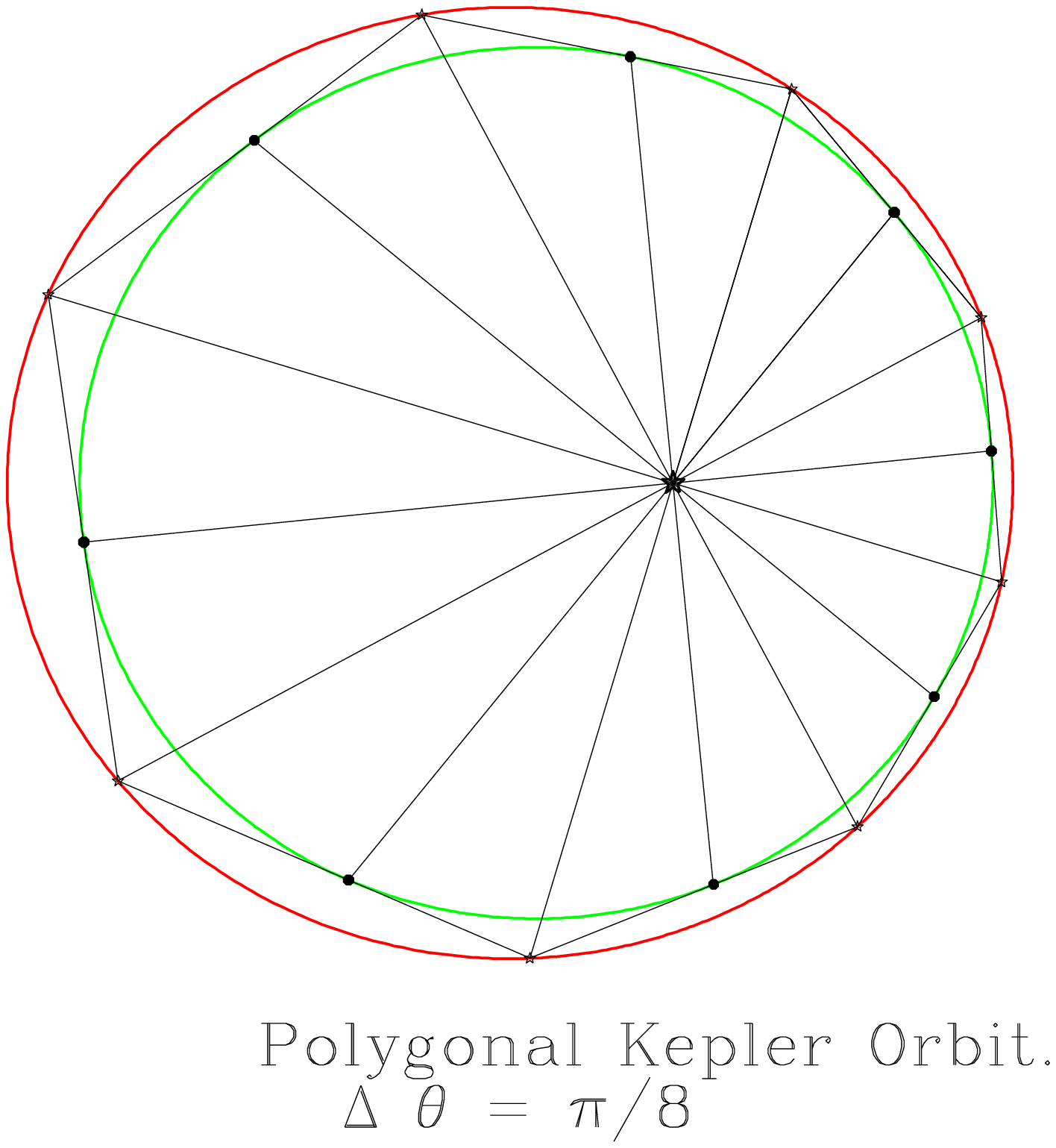}{This is the polygonal approximation to the Kepler orbit with 
	$\Delta\theta= {\pi/ 8}$. The eccentricity of the inner  ellipse
is .3 and of the outer ${.3/ \cos(\pi/8)}=.325$}

\section{Ellipse}
To show that the expression
\bea
r+\epsilon x=R_0
\eea 
where $r^2=x^2+y^2$, is an ellipse we can define
\bea
x'=x+{2\epsilon R_0\over 1-\epsilon^2}\\
r'=x'^2+y^2
\eea
to get 
\bea
0&=&r^2- (-\epsilon x +R_0)^2 =(1-\epsilon^2) x^2 +y^2+2\epsilon xR_0-R_0^2\\
 &=&r'^2 -(\epsilon x'+R_0)^2
\eea
or
\bea
r'-\epsilon x'=R_0
\eea
This gives
\bea
r+r' +\epsilon (x-x')= 2R_0
\eea
or
\bea
r+r'={R_0\over 1-\epsilon^2}
\eea
 $ r$ and $r'$ are the two distances  from the foci of the figure to a point on the ellipse.
 That the sum of the distances of a point on the figure to the two foci
is constant 
is one of the definitions of the ellipse.

If $v_0$ is sufficiently large, then the orbit is no longer an ellipse but
rather a parabola or a hyperbola ($\epsilon=1$ or $\epsilon>1$.)
Thus this proof is valid for all conic sections. 
Furthermore if $\mu<0$ (repulsive force law), $R_0$ and $\epsilon$ are both less than 0, and the
figure is again a hyperbola, but with the centre of force  at the outside
focus of the hyperbola
for any value of $\epsilon<0$.

\end{document}